# Öffentliche Daten auf die nächste Stufe heben

## Vom RESTful Webservice für Pflanzenschutzmittelregistrierungsdaten zur anwendungsunabhängigen Ontologie (erweiterte Version )

Katharina Albrecht [1], Kristoffer Janis Schneider[1], Daniel Martini[1]

**Abstract:** Bei der Ausbringung von Pflanzenschutzmitteln müssen Abstandsauflagen berücksichtigt werden. Allerdings müssen diese händisch ermittelt und berücksichtigt werden. Um den Landwirt dabei zu unterstützen, wurde im PAM-Projekt ein Entscheidungshilfesystem entwickelt. Ein Teil dieses Systems ist der Abstandsauflagendienst, den das KTBL bereitstellt. Für diesen Dienst wurde ein Prozess entwickelt, welcher die Pflanzenschutzmittelregistrierungsdaten aus einer REST-Schnittstelle ausliest, die ausgelesenen Informationen mit Hilfe eines RML Mappings in eine Ontologie überführt und diese über einen SPARQL-Endpoint anwendungsunabhängig abrufbar macht.

## 1 Einleitung

Die maschinelle Ausbringung von Pflanzenschutzmitteln (PSM) ist Stand der Technik; Schläge werden GPS-geführt befahren, mit Applikationskarten können Abstände zu Saumstrukturen, Oberflächengewässern oder urbanen Strukturen definiert werden und über die Teilbreitenabschaltung können die PSM gezielt appliziert werden.

Dies geschieht jedoch nicht automatisch, noch immer obliegt es dem Landwirt diese einzuhaltenden Abstände manuell aus den Auflagentexten in der vom Bundesamt für Verbraucherschutz und Lebensmittelsicherheit (BVL) bereitgestellten Zulassungsdatenbank, im Folgenden PSM-DB genannt, zu ermitteln und diese entweder in die Applikationskarte einzupflegen oder während der Ausbringung einzuhalten. Des Weiteren müssen Aktualisierungen regelmäßig überprüft und vorgenommen werden. Dieses Vorgehen ist nicht nur zeitintensiv, sondern auch fehleranfällig.

Um den Landwirt dabei zu unterstützen, wurde im Pesticide Application Manager (PAM) Projekt und dessen Nachfolgeprojekten PAMrobust, PAM3D und PAM-M ein Entscheidungshilfesystem [Sc16] entwickelt. Während der Projektlaufzeiten ist am

[1] Kuratorium für Technik und Bauwesen in der Landwirtschaft e.V., Digitale Technologien, Bartningstraße 49, 64289 Darmstadt, k.albrecht@ktbl.de, https://orcid.org/0000-0002-3400-8102, k.schneider@ktbl.de, d.martini@ktbl.de

KTBL ein Prozess entstanden, welcher die Auflagendaten in einer Ontologie speichert und mit Augenmerk auf die Nutzung von Standards mit Hilfe von Linked Open Data (LOD) bereitstellt. Das Ergebnis, der Abstandsauflagendienst, ist eine maschinenlesbare und anwendungsunabhängige Repräsentation der durch die PSM-DB bereitgestellten Abstandsauflagen, die somit automatisiert in Applikationskarten übertragen werden können. Dabei werden die Zulassungsdaten mit Hilfe eines deklarativen Mappings aus der vorhandenen relationalen Datenbank in als Resource Description Framework (RDF) [BG04] triples repräsentierte Aussagen überführt. Den Ansatz dafür bildet das in [Be09] erläuterte Vorgehen. Die triples wiederum werden in einem Triplestore bereitgestellt und mit Hilfe von SPARQL Queries abgefragt und über einen Webservice bereitgestellt.

Das BVL hat im Jahr 2020 an der Initiative tech4germany[2] teilgenommen und stellt seitdem die PSM-DB über eine RESTful Schnittstelle bereit. So ist es möglich mit geeigneten technischen Werkzeugen auf die verschiedenen Tabellen zuzugreifen. Dies ist ein bedeutender Schritt hinsichtlich der Maschinenlesbarkeit der Zulassungsdaten.

## 2   Stand des Wissens

In den vergangenen PAM-Projekten wurde der Abstandsauflagendienst entwickelt und fortwährend an neue Entwicklungen angepasst und verbessert. [Ma15] stellt die initiale Entwicklung eines Mappingprozesses aus relationalen Daten in eine Ontologie dar und [Ma18] beschreibt den PSM Mappingprozess, der während des ersten PAM-Projektes entstanden ist. Die Daten werden mittels Mappingbeschreibung aus einer relationalen Datenbank in einen LOD-Service überführt und bereitgestellt. Aufgrund der Vielzahl von Medienbrüchen und Umstrukturierungen wurde im Folgeprojekt PAMrobust besonderer Wert auf die Nutzung von Standards und Reduktion der Medienbrüche gelegt. Dieser weiter entwickelte Mappingprozess ist in [AMS19] zusammengefasst. Die Datenstruktur und inhaltliche Darstellung wurden in PAM3D verfeinert und bilden somit die Basis für die nachfolgend beschriebenen Arbeiten, welche während der Laufzeit von PAM-M durchgeführt wurden.

In Abb. 1 ist der bisherige Mappingprozess dargestellt. Zunächst bildet eine relationale Datenbank, in diesem Fall eine Access Datenbank, die Datengrundlage. Dabei handelt es sich um eine große Anzahl von Tabellen, welche mit Hilfe des Mappingtools db2triples, welches die Mappingsprache R2RML [DSC12] implementiert, in RDF triples übersetzt werden. Diese Triples werden über den Apache Jena Fuseki[3] Triplestore publiziert, sodass über SPARQL-Abfragen [HS13] auf die Daten zugegriffen werden kann. Zusätzlich werden die Informationen über einen Webdienst des KTBL veröffentlicht.

---

[2] https://tech.4germany.org/project/psm/
[3] https://jena.apache.org/documentation/fuseki2/

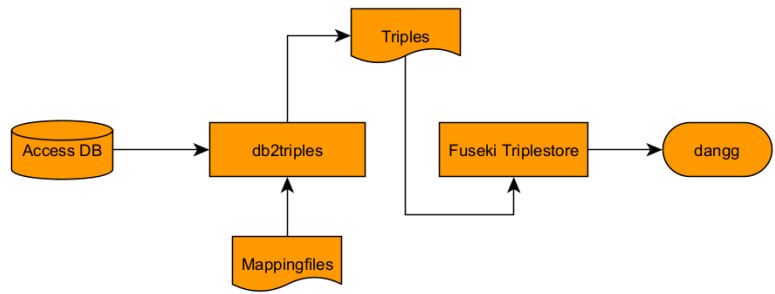

Abb. 1: PSM-Daten liegen in einer Access DB vor, werden mit db2triples gemappt und über den Fuseki Triplestore via Webdienst dangg bereitgestellt

Bisher bestand hinsichtlich der Aktualität der bereitgestellten Ontologie eine starke Abhängigkeit gegenüber der Access DB, welche manuell bereitgestellt, gespeichert und verarbeitet wurde. Dank der Bereitstellung der PSM-DB als REST-Service ist diese Abhängigkeit nicht mehr gegeben. So ermöglicht die RESTful Bereitstellung zukünftig ein vollständig automatisiertes Auslesen der Daten.

Für den PAM-Prozess bedeutet dies zunächst eine Überarbeitung des bestehenden Vorgehens. Es muss ein neuer Prozess zur Abfrage der REST API implementiert werden und das darauf aufbauende Mapping, sowie die Bereitstellung der Daten müssen den neuen Ansprüchen genügen. Zudem wurde das Quellcode-Repositorium für db2triples auf GitHub vom ursprünglichen Bereitsteller geschlossen und das Werkzeug derzeit offenbar nicht mehr weiterentwickelt oder aktualisiert. Daher muss für das Mapping eine Alternative gefunden und eingesetzt werden.

## 3 Der PAM-Prozess

Abb. 2 zeigt den neuen PAM-Prozess. Die PSM-DB, welche über den RESTful Webservice bereitgestellt wird, wird vom PAM-DB-Crawler[4] abgerufen und in einer SQLite Datenbank gespeichert. Dies wird in Abschnitt 3.1 erläutert. Im nachfolgenden Abschnitt 3.2 wird das Mapping erläutert, dabei werden die Informationen aus der SQLite DB mit Hilfe von Mapping Dateien und des Mappingtools RMLMapper[5] von rml.io in RDF triples übertragen. In Abschnitt 3.3 folgt dann die Erläuterung der Bereitstellung per Fuseki Triplestore und des am KTBL entwickelten Webservices.

---

[4] https://github.com/KTBL/PAM-DB-Crawler
[5] https://github.com/RMLio/rmlmapper-java

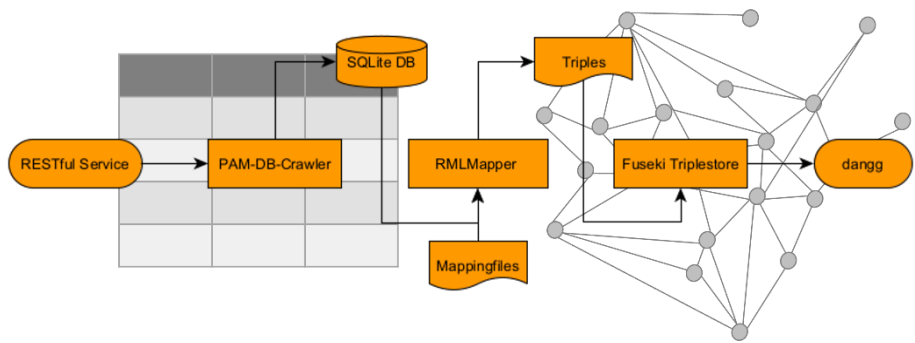

Abb. 2: PSM-Daten werden als RESTful Service bereitgestellt, mit dem PAM-DB-Crawler abgefragt und in einer SQLite DB gespeichert. Das Mapping wird mit dem RMLMapper prozessiert und die Bereitstellung erfolgt über einen Fuseki Triplestore und via Webservice dangg

### 3.1 PAM-DB-Crawler

Die Inhalte der PSM-DB werden vom BVL bereitgestellt und können dort via REST-API abgefragt werden. Die Dokumentation des REST-Service[6] ist mittels Swagger gemäß OpenAPI-Standard zur Verfügung gestellt. Der REST-Service basiert laut Dokumentation auf dem Oracle REST Data Services System, welches die Daten im JSON-Datenformat zurückliefert.

Je Tabelle aus der PSM-DB wird eine Schnittstelle mit dem Format „/<name der Tabelle>/" geboten. Zusätzlich gibt es noch die Schnittstelle „/stand/", welche generelle Informationen über das Release-Datum des aktuellen Datenbestands liefert. Alle Schnittstellen können per HTTP-GET abgefragt werden. Ohne weitere Parameter werden dabei alle verfügbaren Daten zurückgegeben. Je nach Tabelle ist es möglich unterschiedliche Parameter mitzugeben, um nur einzelne Elemente abzufragen oder nach gewissen Kriterien zu filtern. Da die meisten Schnittstellen sehr große Datenmengen zurückliefern (deutlich über 100 Datenelemente) wird ein Paging-Mechanismus genutzt. Das bedeutet, dass die API-Schnittstellen mit jeder Abfrage nur einen Teil des gesamten Ergebnisses zurückgeben. Um weitere Elemente abzufragen kann der „offset" Parameter URL-kodiert mitgegeben werden. Standardmäßig werden so maximal 100 Elemente auf einmal zurückgegeben. Bei Bedarf kann die Page-Größe via des „limit" Parameter verändert werden. Damit bietet sich ein einheitliches System zur Abfrage aller PSM-Daten. Weiter werden auf der Dokumentationsseite auch die je Schnittstelle genutzten Datenmodelle definiert. Da der REST-Service auf einer Oracle Datenbank basiert, sind die genutzten Datentypen als Oracle-Datentypen deklariert.

---

[6] https://psm-api.bvl.bund.de

Um das für die Überführung aus der früher genutzten Access-Datenbank bereits entwickelte R2RML-Mapping weiter nutzen zu können, werden die PSM-Daten in einer relationalen Datenbank benötigt. Mithilfe des REST-Service bietet sich die Möglichkeit alle Daten abzufragen und in einer lokalen Datenbank zu speichern. Von dort aus können sie mittels R2RML-Mapping in eine Ontologie überführt werden. Ein Weg dies umzusetzen wäre die Datenmodelldefinitionen zu nutzen, um händisch passende Datenstrukturen zu erstellen, diese mittels API-Abfragen zu befüllen und die abgefragten Daten in einer SQLite Datenbank zu speichern. Damit umfasst eine Implementierung die folgenden Schritte je abgefragter Tabelle:

- Implementieren der Datenstrukturen
- Implementieren der API-Abfragen, um jeweils die Datenstrukturen zu befüllen
- SQL-Statements zur Erzeugung der SQLite Tabellen
- SQL-Insert-Statements zum Befüllen der SQLite Tabellen

Bei diesem Vorgehen ist viel Boilerplate-Code notwendig, da zwar alle Datenstrukturen und APIs gleichartig aufgebaut und genutzt werden können, aber aufgrund der inhaltlichen Unterschiede alle Elemente getrennt implementiert werden. Zur Reduktion des Boilerplate-Codes bietet es sich an aufgrund eben dieser Gleichartigkeit anstatt Strukturen, SQL-Statements und Abfragen manuell zu implementieren, einen Generator im Sinne der Meta-Programmierung zu verwenden. Ein solcher Generator erzeugt auf Basis von Definitionsinformationen Code, welcher alle oben skizzierten Schritte von der Struktur bis hin zu den Insert-Statements umfasst.

Für dieses Vorgehen wurde ein einfacher Generator in Python implementiert, welcher wiederum Python-Code als Ergebnis liefert. Als Definitionsinformationen dient die Kombination aus API-Endpoint-Adresse und Modelldefinition (aus der Service-Dokumentation). Daraus werden sowohl Datenstrukturen als auch Create- und Insert-Statements, sowie Code zum Abfragen der API und Speichern der geladenen Daten generiert. Als Grundlage für den Generator dienen Format-Strings welche Python-Code als Text beinhalten und an entsprechenden Stellen Struktur- bzw. API-spezifische Informationen eingesetzt bekommen. Typinformationen für die SQLite-Tabellen werden aus den Modelldefinitionen der Dokumentation übernommen. Da die Typen aus Oracle sich von den SQLite-Typen unterscheiden, besitzt das Projekt ein Mapping zwischen beiden Typ-Welten. Der erzeugte Python-Code wird in Text-Dateien mit der Dateiendung „.py" gespeichert. Ein weiteres Script, welches nicht generiert wird, dient als ausführendes Script um den Download zu managen. Dafür lädt dieses Script alle dynamisch erzeugten Strukturdefinitionen, initialisiert falls nötig die Datenbank und startet anschließend den Ladeprozess der Daten. Falls der Prozess unterbrochen wird, wird der aktuelle Stand abgespeichert und er kann zu einem späteren Zeitpunkt wieder aufgenommen werden.

Ein großer Vorteil dieses Vorgehens ist, dass Änderungen am Datenmodell oder an der Adresse der Schnittstellen schnell eingepflegt und auf verschiedenen Wegen angepasst werden können. Ein Weg ist die Datendefinitionen zu ändern und anschließend den

Generator zu starten, um den Python-Code neu zu generieren. Die andere Möglichkeit ist, händisch den bereits generierten Python-Code anzupassen, da dieser als Code-Text vorliegt. In letzterem Fall muss man jedoch beachten, dass ein erneutes Starten des Generators Änderungen potenziell überschreibt.

### 3.2 PSM Mapping

Die nach Ausführung des PAM-DB-Crawlers in einer SQLite DB vorliegenden PSM-Daten werden mit Hilfe eines Mappingverfahrens aus der relationalen Darstellung in RDF triple bzw. die Ontologie überführt. Allerdings ist das bisher genutzte Mappingtool db2triples auf Github nicht mehr verfügbar. Da somit keine Aktualisierungen oder Bugfixes für das Tool vorhanden sind, ist die Nutzung einer Alternative notwendig. Nach eingehender Recherche, welche die Standardkonformität, sowie die freie Verfügbarkeit berücksichtigte, wurde der RMLMapper-java von rml.io als passendes Tool identifiziert. Der RMLMapper ist eine Java Bibliothek, welche mit Hilfe von RML-Regeln [DHD20] Linked Data generiert. RML basiert auf R2RML und erweitert dieses. Der RMLMapper unterstützt eine Reihe von sowohl lokalen als auch remote vorliegenden Datenquellen, welche als Inputdaten für das Mapping genutzt werden können. Allerdings sind hier weder SQLite noch Access-Datenbanken berücksichtigt. Um die Anbindung zu realisieren, wurde das Github Repository des RMLMappers geklont und die spezifische Datenbankanbindung in den Quellcode integriert.

Um das Mapping mit rmlmapper-java zu testen, wurde zunächst die vorhandene Access DB als Input genutzt. Das Ergebnis entsprach den Erwartungen, sodass die SQLite Anbindung umgesetzt werden konnte. Im Folgenden werden die einzelnen Schritte des Mappings erläutert.

**Relationale Darstellung**

Abb. 3 zeigt Auszüge aus zwei Tabellen der PSM-DB, wie sie sowohl in der Access DB als auch in der SQLite DB vorliegen. Tabelle AWG beinhaltet übergreifende Informationen zu den jeweiligen Indikationen und bildet während des Mappings den Ankerpunkt, über den später alle Informationen abgerufen werden. Dabei ist die AWG_ID die ID der Indikation. Die zweite Tabelle zeigt einen Auszug aus AWG_KULTUR, also jene Kultur, auf die das Pflanzenschutzmittel in der jeweiligen Indikation angewandt wird. Diese Tabelle ist eine Verknüpfungstabelle, welche die Kulturen der Indikation zuordnet. Um aus diesen zwei Tabellen eine Aussage, also ein RDF triple zu erstellen, müssen diese gemappt werden. Dafür wird nachfolgendes Mapping verwendet.

| AWG_ID | ... |
|---|---|
| 034028-60/07-004 | ... |
| 034028-60/07-006 | ... |
| ... | ... |

| AWG_ID | KULTUR | ... |
|---|---|---|
| 034028-60/07-004 | RUBID | ... |
| 034028-60/07-006 | RUBFR | ... |
| ... | ... | ... |

Abb. 3 links: ein Auszug aus Tabelle AWG, rechts: ein Auszug aus Tabelle AWG_KULTUR

**Identifizierbarkeit**

Ziel des Mappings ist die Erstellung einer PSM Ontologie, welche alle notwendigen Informationen zu Indikationen, Mitteln, Kulturen, Schaderregern und Abstandsauflagen enthält. Dafür müssen die einzelnen Informationen eindeutig zuzuordnen sein.

Da im Bereich Linked Data die Identifier zudem auflösbar sein sollen[7], wie auch das FAIR Principle A1[8] beschreibt, sind nicht auflösbare URNs (Unified Resource Names) nicht nutzbar. Daher werden für Ontologien in der Regel URIs (Unified Resource Identifier) genutzt. Diese URIs ermöglichen eine eindeutige Identifizierbarkeit der einzelnen Ressourcen, die während des Mappings definiert werden. Die Eindeutigkeit ermöglicht es zudem, dass Informationen aus verschiedenen Quellen oder Mappings richtig zusammengesetzt und beliebig erweitert werden können. Andersherum können Informationen eindeutig abgefragt und ermittelt werden.

Für die PSM Ontologie wurde das bis dahin genutzte URI-System überarbeitet. URIs werden nun deutlicher in Beziehung zu ihren Inhalten gesetzt, also vorliegenden Beispiel der Indikation und der Kultur. Das Vokabular der Ontologie, also die notwendigen Classes und Properties erhalten den Pfad *http://srv.ktbl.de/data/psm/* und die konkreten Instanzen sind über *http://srv.ktbl.de/data/psm/resources/* definiert, wobei dann je nach Klasse zusätzlich ein sogenannter Klassenpräfix hinzugefügt wird. Im gegebenen Beispiel ist das *ind_* für die AWG_ID bzw. Indikation und *crop_* für die Kultur, auf die das Mittel angewendet wird. Ein Teil der URIs ist schon jetzt auflösbar.

---

[7] https://www.w3.org/wiki/LinkedData
[8] https://www.go-fair.org/fair-principles/metadata-retrievable-identifier-standardised-communication-protocol/

### Das Mappingfile

```
@prefix rr: <http://www.w3.org/ns/r2rml#>.
@prefix rml: <http://semweb.mmlab.be/ns/rml#>.
@prefix psm: <http://srv.ktbl.de/data/psm/>.
@prefix ql: <http://semweb.mmlab.be/ns/ql#>.

@prefix d2rq:<http://www.wiwiss.fu-berlin.de/suhl/bizer/D2RQ/0.1#> .
<#DB_source> a d2rq:Database;
    d2rq:jdbcDSN "jdbc:sqlite://Pfad/Name;
    d2rq:jdbcDriver "org.sqlite.JDBC";
    d2rq:username "";
    d2rq:password "".

<#TriplesMap1> a rr:TriplesMap;
    rml:logicalSource [
        rml:source <#DB_source> ;
        rr:sqlVersion rr:SQL2008;
        rml:referenceFormulation ql:CSV;
        rml:query """ SELECT AWG_ID, REPLACE(AWG_ID, "/", "-") AS URLID FROM AWG
        """
    ];
    rr:subjectMap [
        rr:template "http://srv.ktbl.de/data/psm/resources/ind_{URLID}";
        rr:class psm:Indication;
    ];
    rr:predicateObjectMap [
        rr:predicate psm:appliedOnCrop;
        rr:objectMap [
            rr:parentTriplesMap <#TriplesMap2>;
            rr:joinCondition [
                rr:child "AWG_ID";
                rr:parent "AWG_ID";
            ];
        ];
    ].
    <#TriplesMap2> a rr:TriplesMap;
    rml:logicalSource [
        rml:source <#DB_source> ;
        rr:sqlVersion rr:SQL2008;
        rml:referenceFormulation ql:CSV;
        rml:query """ SELECT AWG_ID, KULTUR FROM AWG_KULTUR """
    ];
    rr:subjectMap [
        rr:template "http://srv.ktbl.de/data/psm/resources/crop_{KULTUR}";
        rr:class psm:Crop;
    ].
```

Abb. 4: Beispiel für eine Mappingdatei: Auszug aus Map_Crop.ttl

Ein Mapping für das oben dargestellte Beispiel wird in Abb. 4 mit Hilfe eines Auszugs aus der Map_Crop.ttl gezeigt. Die Inhalte der Mappingdatei, welche mit R2RML bzw. RML beschrieben werden, bilden selbst einen RDF Graph in der Syntax Turtle [PC13]. Die vollständigen Mappingfiles der PSM-DB finden sich im Github Repositorium RML-Mapping[9] des KTBL.

Alle Mapping-Dateien besitzen einen Kopf, welcher alle in der Datei genutzten Präfixe enthält. Diese Präfixe ermöglichen eine kompakte und gut lesbare Darstellung der Inhalte. Im Beispiel werden die von RML vorgegebenen Präfixe *rr*, *rml, ql* und *d2rq* verwendet, sowie das Präfix *psm*, welches für alle Inhalte, welche mit den PSM in Zusammenhang stehen genutzt wird.

Danach folgt der deklarative Teil, welcher die sogenannten Triples Maps enthält. Triples Maps können einzeln und unabhängig definiert werden, oder aber in Beziehung stehen. Dafür wird eine Eltern-Kind-Beziehung mit Hilfe einer *JoinCondition* definiert.

Da RML auf R2RML aufbaut, ist der Unterschied in den Mapping Dateien im Allgemeinen nur gering. Im Gegensatz jedoch zu R2RML Mappings, die nur auf relationale Datenbanken angewendet werden können, sind bei RML verschiedene Datenformate als Inputdatei möglich. Der „Source Header" wird daher benötigt, um angeben zu können, um welche Art von Inputdatei es sich handelt. Im vorliegenden Beispiel beschreibt *<#DB_source>* die erstellte SQLite DB. Zusätzlich wird der passende JDBC Treiber angegeben. Username und Passwort sind bei dieser Datenbank nicht notwendig.

*<#TriplesMap1>* beschreibt das erste Mapping. Die *rml:logicalSource* enthält die Informationen zur genutzten Inputdatei, der SQL-Version (SQL 2008) und der Reference Formulation (CSV). Da auf eine Datenbank zugegriffen wird, ist es zusätzlich möglich über *rml:query* pro Triples Map eine SQL Query definiert werden. Während des Mappings müssen zudem Feinheiten der SQL-Abfragesprachen von Access und SQLite berücksichtigt werden. Da einige Abstandsauflagen nur in Auflagentexten vorliegen, ist die Nutzung der SQL Query unumgänglich. So werden konkrete Inhalte aus den Auflagentexten beispielsweise mit Hilfe von regulären Ausdrücken (RegEx) extrahiert. Dies ermöglicht es, die PSM-Daten so aufzubereiten, dass die einzuhaltenden Abstände maschineninterpretierbar, also nur als Zahl, getrennt von ihrer Einheit, bereitgestellt werden können.

Innerhalb einer Triples Map muss immer eine SubjectMap und mindestens eine PredicateObjektMap definiert werden. Die SubjectMap repräsentiert, wie der Name schon sagt, das Subjekt des RDF triples. Im Beispiel ist das Subjekt die Indikation. Da der Slash („/") in der Zeichenfolge der AWG_ID ein reserviertes Zeichen innerhalb des URL-Encodings ist, wird dieser innerhalb der SQL Query durch einen Bindestrich ersetzt. Über *template* wird die URI des Subjekts festgelegt und mit *rr:Class* wird dem Subjekt die Klasse *psm:Indication* zugewiesen.

---

[9] https://github.com/KTBL/RML-Mapping

Nach der SubjectMap wird eine PredicateObjectMap definiert. Ein Objekt kann sowohl eine Ressource, ein Literal oder ein Blank Node sein. Über das Prädikat *psm:appliedOnCrop* wird ausgedrückt, dass das genutzte PSM in dieser Indikation auf eine bestimmte Kultur angewandt wird. Dafür wird eine *Join Condition* genutzt, welche in beiden Tabellen nach dem gemeinsamen Eintrag der angegebenen Tabellenspalte (AWG_ID) sucht.

<#TriplesMap2> zeigt die Erstellung der Kultur. Die *rml:logicalSource* ist analog zu <#TriplesMap1> definiert. Der Zugriff auf die Kultur-Tabelle wird wieder in der SQL Query definiert. Es muss sowohl die Spalte der Kultur als auch die AWG_ID ausgelesen werden, da sonst die *Join Condition* die Beziehung der Triples Maps nicht herstellen kann.

**Das Mappingergebnis**

Mit `java -jar rmlmapper-VERSION-all.jar -m Map_Crop.ttl -o Res_Crop.ttl -s turtle` werden das Mapping durchgeführt und die Ergebnistriple in der Datei *Res_Crop.ttl* im Turtle-Format gespeichert. Diese Dateien werden dann in den Triplestore geladen. Für das obige Beispiel ist der Mappingoutput folgender:

```
@prefix psm: <http://srv.ktbl.de/data/psm/> .

@prefix psmr: <http://srv.ktbl.de/data/psm/resources/> .

psmr:ind_034028-60-07-004 a psm:Indication;
   psm:appliedOnCrop psmr:crop_RUBID .

psmr:ind_034028-60-07-006 a psm:Indication;
   psm:appliedOnCrop psmr:crop_RUBFR .

psmr:crop_RUBFR a psm:Crop.

psmr:crop_RUBID a psm:Crop.
```

Die einzelnen Zeilen bilden logische Aussagen der Form Subjekt-Prädikat-Objekt. So ist zu sehen, dass psmr:ind_034028-60-07-004 eine Instanz der Klasse psm:Indication ist. Das Semikolon zeigt an, dass die darauffolgende Zeile das gleiche Subjekt besitzt, also dass psmr:ind_034028-60-07-004 auf die Kultur psmr:crop_RUBID angewendet wird. Die gleiche Aussage wird für psmr:crop_RUBFR getroffen. psmr:crop_RUBID und psmr:crop_RUBFR wiederum sind Instanzen der Klasse psm:Crop.

### 3.3 Triplestore und Bereitstellung

Um die generierte Ontologie erreichbar zu machen, müssen die triple in einen Triplestore geladen werden. Dieser Triplestore wird über einen SPARQL-Endpoint verfügbar gemacht und kann dann mittels der standardisierten Abfragesprache SPARQL abgefragt werden.

Zur Publikation der Daten wird ein Apache Jena Fuseki Server genutzt. Der Fuseki kann sowohl auf dem Desktop PC, auf Virtuellen Maschinen (VM) oder in einem Docker[10] Container laufen. Letzteres spielt vor allem für die mögliche Portabilität des Systems eine wichtige Rolle, da Docker Container von der eigentlichen Hardware des Hauptsystems unabhängig sind.

SPARQL Queries können entweder direkt in der Benutzeroberfläche des Fuseki oder per HTTP Request abgesetzt werden. Abb. 5 zeigt ein Beispiel innerhalb der Benutzeroberfläche. Im Beispiel wird eine DESCRIBE Query abgefragt: `DESCRIBE psmr:ind_034028-60-07-004`. Mit dieser Query werden alle Informationen zurückgegeben, die mit der gesuchten Indikation in direkter Verbindung stehen; also die Aussage, dass die gesuchte Ressource eine Indikation ist und dass diese Indikation auf die Kultur RUBID angewandt wird. Neben der Antwort auf die Query sind auch alle Präfixe aufgeführt, welche in der Ontologie vorhanden sind.

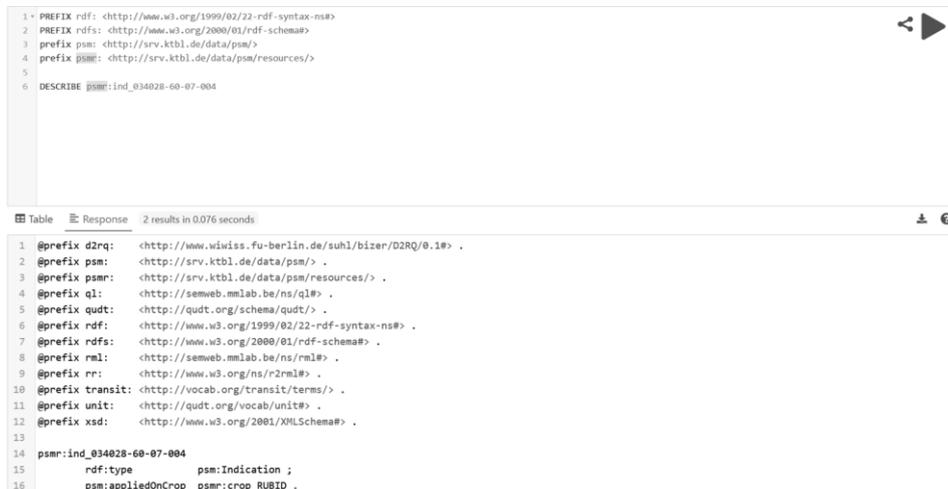

Abb. 5 : Beispiel Rückgabe nach Describe Query in der Fuseki Oberfläche

Weiterhin können SELECT Queries definiert werden, welche einfache Abfragen mit tabellarischer Ausgabe ähnlich SQL ermöglichen, oder CONSTRUCT Queries, mit deren Hilfe selektiv Untergraphen erstellt werden können. Sofern bei der Implementierung des Endpoints vorgesehen, sind auch UPDATE Queries möglich. Als Rückgabeformat wird im Beispiel Turtle genutzt, es sind jedoch eine Reihe weiterer Formate, darunter JSON-LD, N-Triples oder XML, möglich.

---

[10] https://www.docker.com/

## 4  Ergebnisse und Diskussion

Für den Abstandsauflagendienst wurde ein eigener Webservice in go[11], genannt „dangg", implementiert. Dieser gibt unter anderem alle Abstandsauflagen, welche für eine bestimmte Indikation berücksichtigt werden müssen, im JSON-LD[12] Format zurück. Ein Auszug aus der Rückgabe ist in Abb. 6 dargestellt. Der Auszug zeigt die Indikation 034028-60/07-004 mit der zugehörigen Kultur, dem Schaderreger, dem Mittel, der Applikationszeit, der Tankmischung, den Auflagen und den Abstandsauflagen. Die Verweise auf beispielsweise _:b4 stehen für Blank Nodes, welche weitere Inhalte bereitstellen. Diese Darstellung wird von Fuseki geliefert. Die Key-Value Paare der Rückgabe enthalten die eindeutigen Identifier der einzelnen Ressourcen, sodass man nach dem vollständigen Mapping weitere Informationen zu Mitteln oder Kultur/Schaderreger abfragen kann.

```
}, {
    "@id" : "psmr:ind_034028-60-07-004",
    "@type" : "psm:Indication",
    "agent" : "psmr:ex_034028-60",
    "applicationTime" : "psmr:at_BF",
    "appliedOnCrop" : "psmr:crop_RUBID",
    "appliedOnPest" : "psmr:pest_PHRARU",
    "considerBufferConstraint" : "_:b14",
    "constraint" : [ "psmr:cons_NW605", "psmr:cons_NT102", "psmr:cons_WW7091", "psmr:cons_NW606", "psmr:cons_NW701" ],
    "scopeOfApplication" : "psmr:foa_O",
    "usableApplicationRate" : "_:b6",
    "label" : "034028-60/07-004"
} ],
```

Abb. 6: Auszug aus JSON-LD Rückgabe für Beispielindikation

Die Bereitstellung der PSM-DB über eine RESTful Schnittstelle vereinfacht den PAM Prozess deutlich. Durch die Nutzung des PAM-DB-Crawlers, können die Daten flexibel in die SQLite DB gespeichert und gemappt werden. Der RMLMapper ist ein adäquater Ersatz für db2triples und bietet Raum für weitere Verbesserungen z.B. Einbeziehen von weiteren Datenquellen in anderen Formaten in das Mapping. Die Nutzung von Fuseki hat sich ebenfalls bewährt. Insgesamt bietet der PAM Prozess nun einen zuverlässigen Weg um die Pflanzenschutzmittelregistrierungsdaten anwendungsunabhängig und maschinenlesbar bereitzustellen.

---

[11] https://go.dev/
[12] https://json-ld.org/

## 5    Ausblick

Die BVL API ermöglicht die vollständige Automatisierung des PAM Prozesses. Die Entwicklung des PAM-DB-Crawlers ist dabei ein essenzieller Schritt hinsichtlich dieser Automatisierung. Mit Hilfe eines Batch-Jobs kann zukünftig implementiert werden, dass der PAM-DB-Crawler die monatlichen Updates der PSM-DB abruft, um ein zeitnahes Mapping zu ermöglichen. Um Redundanzen und Fehlinformationen zu vermeiden, muss hier jedoch noch ein geeignetes Vorgehen entwickelt werden. Der RMLMapper ermöglicht das Speichern der triples in verschiedenen Formaten. Dazu gehören neben dem gut menschenlesbaren turtle Format unter anderem auch ntriples[13] und nquads[14]. Diese Darstellungen können bei der Versionierung helfen, da sie zeilenweises Vergleichen von Aussagen ermöglichen.

Des Weiteren ist es mit dem RMLMapper möglich, das erstellte Mapping automatisch auf einen SPARQL Endpoint via SPARQL-Update zu publizieren, sodass in Zukunft der Schritt des manuellen Hochladens entfallen kann. Hier sind jedoch noch Überlegungen zum Aufbau einer stabilen Infrastruktur notwendig, um sowohl eine stabile Bereitstellung als auch einen sicheren Betrieb gewährleisten zu können.

Die aktuell genutzte Version des Fuseki ist 4.3.2.. Beim nächsten Update wird es möglich sein, in der JSON-LD Darstellung die Keys mit Präfixen zurückzugeben, sodass hier eine stärkere und einfachere Kopplung zur eigentlichen Ontologie möglich ist.

Um möglichst unabhängig von verschiedenen Infrastrukturen zu sein, soll der ganze Prozess mit Hilfe von Docker umgesetzt werden. Dies ermöglicht zudem einfachen Austausch bzw. einfache Weitergabe der Prozessmodule für weitere Nutzer.

---

[13] https://www.w3.org/TR/n-triples/
[14] https://www.w3.org/TR/n-quads/